# Electron beam irradiation-induced transport and recombination in p-type Gallium Oxide grown on (001) *β*-Ga$_2$O$_3$ substrate

Gabriel Marciaga [1], Jian-Sian Li [2], Chao-Ching Chiang [2], Fan Ren [2], Stephen J. Pearton [3],

Corinne Sartel [4], Zeyu Chi [4], Yves Dumont [4],

Ekaterine Chikoidze [4], Alfons Schulte [1], Arie Ruzin [5], Leonid Chernyak [1, a)]

[1] *Department of Physics, University of Central Florida, Orlando, FL 32816, USA*

[2] *Department of Chemical Engineering, University of Florida, Gainesville, FL 32611, USA*

[3] *Department of Materials Science and Engineering, University of Florida, Gainesville, FL 32611, USA*

[4] *Groupe d'Etude de la Matière Condensée, Université Paris-Saclay, Université de Versailles Saint Quentin en Yvelines – CNRS, 45 Av. des Etats-Unis, 78035 Versailles Cedex, France*

[5] *School of Electrical Engineering, Tel Aviv University, Tel Aviv 69978, Israel*

**Abstract**

This study investigates minority electron diffusion length and carrier recombination phenomena within p-type, 300 nm-thick Ga$_2$O$_3$ epitaxial films. Utilizing Electron Beam-Induced Current (EBIC) and Cathodoluminescence (CL) spectroscopy, these characteristics were examined as a function of both temperature and the duration of electron beam excitation. While the electron diffusion length in these p-Ga$_2$O$_3$ films diminish with increasing temperature, a continuous electron beam excitation of a particular location on the surface of a p-Ga$_2$O$_3$ epitaxial layer leads to an elongation of the diffusion length. and decay of cathodoluminescence intensity in that location under beam exposure. These latter two effects are attributed to non-equilibrium electrons, generated by an electron beam, being captured at acceptor-related point defect levels in Gallium

Oxide. The activation energies characterizing these processes were obtained from the independent EBIC and CL experiments to garner insight into defect landscape and its influence on transport and recombination dynamics.

[a]  Email: leonid.chernyak@ucf.edu

**I. Introduction**

Electron beam excitation significantly impacts the diffusion length and cathodoluminescence (CL) in Gallium Oxide ($Ga_2O_3$), a wide-bandgap semiconductor with promising applications in optoelectronics and power electronics. The effects are primarily attributed to generation (under the influence of electron beam) of non-equilibrium carriers in the material and their interaction with defects within the material [1].

Electron beam irradiation generally leads to a significant elongation of minority carrier diffusion length ($L$) in both n-type and p-type $Ga_2O_3$ [1,2]. This increase often follows a linear trend with the duration of electron beam irradiation before eventually saturating [3].

The primary mechanism for the elongation of $L$ under electron irradiation is attributed to the trapping of non-equilibrium electrons (generated by the primary electron beam) by a high concentration of neutral gallium vacancies at metastable deep acceptor levels within the bandgap of $Ga_2O_3$ [1]. By occupying these defect levels, the injected electrons effectively passivate the defects, preventing them from acting as recombination centers for carriers and thus an increase in carrier lifetime ($\tau$) [1].

With fewer recombination centers available, minority carriers experience a longer $\tau$ and can travel a greater average distance before recombining with majority carriers. The diffusion length is related to the diffusion coefficient ($D$) and the carrier lifetime by the equation [1,4]:

$$L = \sqrt{D\tau} \qquad (1)$$

Therefore, an increase in carrier lifetime directly translates to a longer diffusion length [1,3].

Interestingly, electron beam injection has also been shown to recover the diffusion length in $Ga_2O_3$ samples that have been previously degraded by radiation (e.g., alpha and proton irradiation) [2]. In some cases, the diffusion length after electron injection even surpasses the pre-irradiation values, indicating a "healing" effect of the electron beam.

The rate of increase in diffusion length with electron irradiation and *L* saturation behavior are temperature-dependent, suggesting that the trapping and de-trapping processes at the defect levels are thermally activated. Activation energies for these processes have been estimated to be in the range of 72 meV to 304 meV, via EBIC and CL techniques, respectively. These are linked to Gallium vacancy-related defects [3,5].

Studies have shown that the intensity of CL emission bands can decrease with increasing duration of electron beam irradiation [5]. This decay is linked to the trapping of non-equilibrium electrons at defect levels, which are also involved in radiative recombination processes. By trapping these electrons, the availability of these levels for radiative transitions is reduced, leading to a decrease in CL intensity [6,15].

Prior investigations studied the electron irradiation-induced effects on undoped p-type Gallium Oxide were conducted on epitaxial layers grown on (010)-oriented insulating Fe-doped $Ga_2O_3$ substrates [3,5]. In contrast, this present work will examine minority carrier diffusion length and recombination phenomena using two complementary techniques - Electron Beam-Induced Current (EBIC) and cathodoluminescence (CL) - in undoped p-type $Ga_2O_3$ thin films deposited on conductive (001)-oriented Gallium Oxide substrate doped with tin (Sn). This may highlight similarities and differences in minority carrier transport and recombination dynamics on epitaxially grown p-type $Ga_2O_3$.

## II. Experimental

The β-Ga$_2$O$_3$ epilayer was grown on Sn doped β-Ga$_2$O$_3$ (001) substrates purchased from Novel Crystal Technology, by Metal-Organic Chemical Vapor Deposition (MOCVD). Trimethylgallium (TMGa) as a source for gallium and high-purity oxygen gas were used as precursors. The flow rates of TMGa and O$_2$ were set at 34 μmol/min and 3200 sccm, respectively. The pressure and temperature in the growth chamber were maintained at 40 Torr and 825 °C. The acceptor density in the epilayer between 40 and 200 nm from the top epilayer surface was estimated to be ~ 2×10$^{17}$ cm$^{-3}$ by Capacitance-Voltage (C–V) measurements (cf. Fig. 1).

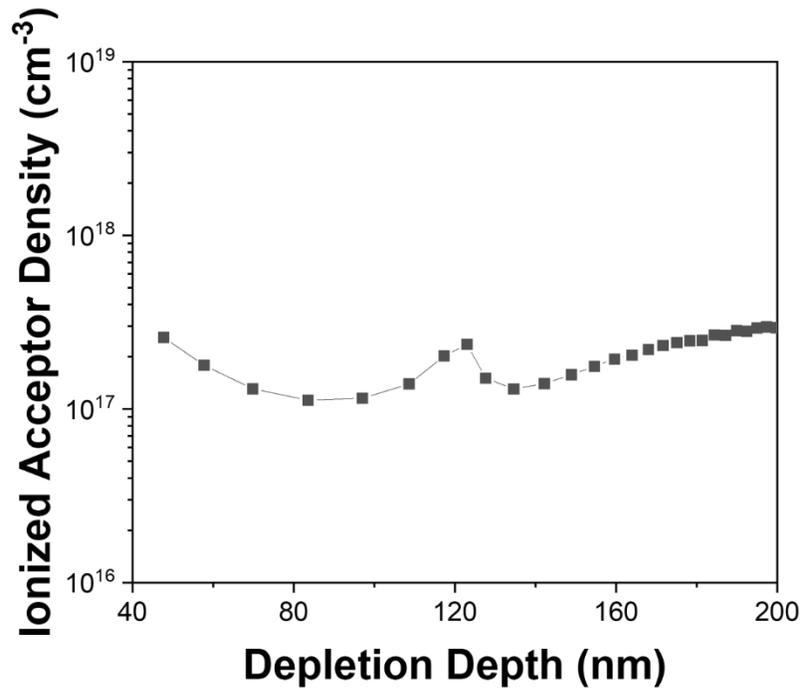

**Fig. 1.** Shows acceptor density in the p-type epilayer between 40 and 200 nm of depth from the surface. The acceptor density was estimated to be ~ 2×10$^{17}$ cm$^{-3}$ by Capacitance-Voltage (C–V) measurements.

Cathodoluminescence (CL) characterizations were conducted across a 50 to 120 °C temperature range with the electron beam operated at an acceleration voltage of 20 kV utilizing a Gatan MonoCL2 attachment on the SEM. The temperature-dependent cathodoluminescence spectra were obtained utilizing a temperature-controlled stage linked to an external control unit. We detected the emitted light with a Hamamatsu photomultiplier tube (sensitive over 150–850 nm) following dispersion by a single-grating monochromator (1200 lines/mm blaze) [6].

Electron Beam-Induced Current (EBIC) measurements performed *in-situ* with a Phillips XL-30 SEM were used to determine minority carrier diffusion length ($L$). This technique utilizes planar line-scan electron beam excitation, with the beam traversing the sample's surface [1,2,7-9]. EBIC characterizations were conducted across 25 to 120 °C temperature range, and an electron beam accelerating voltage of 25 kV. Chosen to ensure full penetration of the epitaxial p-type layer. The corresponding ~ 0.6 nA absorbed current was measured with a Keithley 480 pico-ammeter and the electron interaction volume is estimated to approximately 1 µm penetration depth into the material [10]. For the diffusion length derivation, 16.3 µm lateral EBIC line-scans were executed off the edge of asymmetrical pseudo-Schottky contacts made of Ni/Au (20 nm/80 nm) created on the p-type layer with standard lithography and liftoff techniques. It should be noted that although the thickness of the p-$Ga_2O_3$ epitaxial film under investigation is about 300 nm and the electron beam penetration depth is about 1000 nm, the diffusion length measurements are still reliable. Both contacts used for the measurements are located on the top surface of the structure ensuring the current path is laterally constrained to the thin epitaxial layer.

Each individual line-scan requires ~ 12 s to be completed, which is adequate for determining $L$. The attenuation of the EBIC signal is expressed with the following relationship [11-13]:

$$I(d) = I_o d^\alpha \exp(-\frac{d}{L}) \qquad (2)$$

In the equation above, *I(d)* is the EBIC signal at a given position, *d*; the variable *d* is the distance measured outwards from the Ni/Au contact edge. $I_0$ denotes a scaling factor. *α*, a coefficient for recombination is set to -0.5 for this experiment, and *L* is the minority carrier (electrons) diffusion length.

**III. Results and discussion**

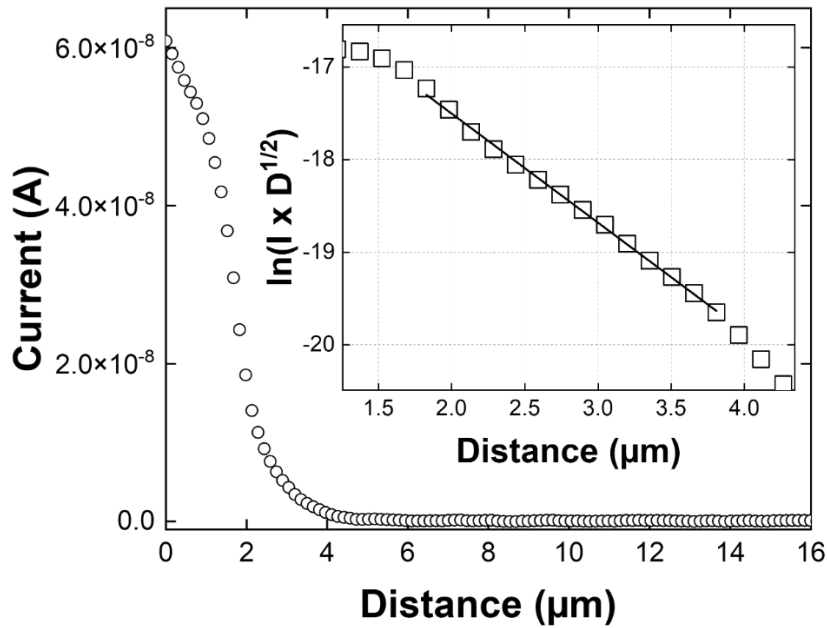

**Fig. 2.** Room temperature characteristics of the initial EBIC signal decrease as function of distance from the Schottky contact's edge. Inset displays $\ln(I \times d^{1/2})$ vs *d*, in accordance with Eq. (2), the minority carrier diffusion length, *L*, is calculated from the negative reciprocal of the linear fit's slope.

Fig. 2 presents the initial, pre-irradiation spatial dependence of the EBIC signal at ambient temperature illustrating the exponential decay with increasing distance from the Schottky edge. The inset plots $\ln(I \times d^{1/2})$ vs *d* for the derivation of *L*, according to Eq. (2). At room temperature the negative reciprocal slope of the line fit yields *L* = 0.85 µm. To investigate the impact of sustained carrier injection, the EBIC measurement area was exposed to continuous electron beam irradiation, accumulating to a total time of 1200 s. During the prolonged period of irradiation, diffusion length was periodically extracted using Eq. (2) at 100 s intervals up to the total 1200 s. It is important to note that at each temperature, the measurement was done at different pristine regions near the contact to avoid unintentional effects from previous tests.

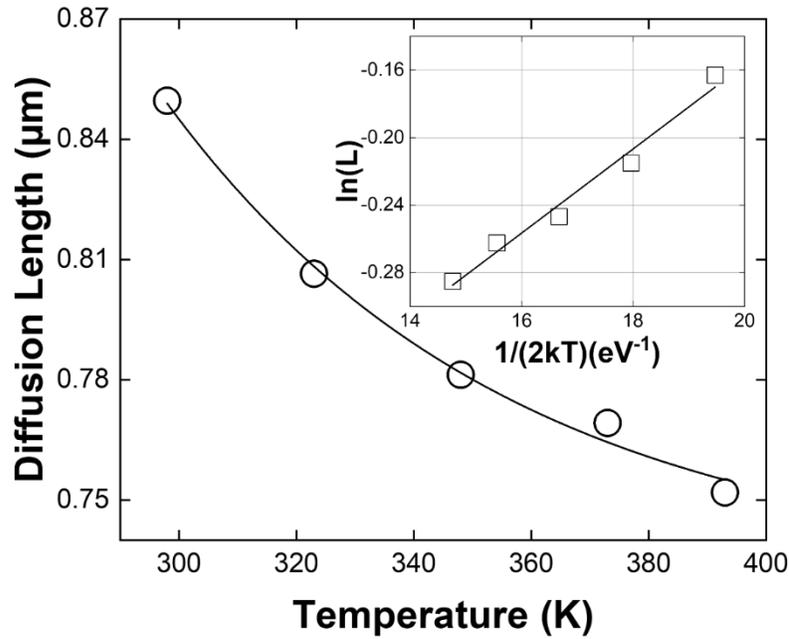

**Fig. 3.** Minority carrier diffusion length, $L$, vs temperature over a 25 to 120 °C range. The data demonstrates the expected exponential trend consistent with Eq. (3). Inset presents an Arrhenius plot of ln ($L$) vs $1/(2kT)$ in order to derive activation energy, $\Delta E_{A,T}$, of 24 meV via Eq (3).

Fig. 3 illustrates the relationship between $L$ and temperature ($T$), which reveals a decrease in $L$ with an increasing temperature. This was achieved via initial $L$ measurements before continuous irradiation onto the sample. This agrees with previous reports on $Ga_2O_3$ [2, 12, 13], where such behavior is ascribed to effects of phonon scattering. The thermal evolution of $L$ is modeled by [9,12]:

$$L(T) = L_0 \, exp\left(\frac{\Delta E_{A,T}}{2kT}\right) \tag{3}$$

In this equation, $L_0$ is a scaling constant; $\Delta E_{A,T}$ is the associated thermal activation energy for this process; $k$ is the Boltzmann constant; and $T$ is temperature. The value of $\Delta E_{A,T}$ can be extracted by linearizing Eq. (3) through a plot of ln (L) vs 1/2kT as seen in the inset of Fig. 3. This activation energy, characterizing the temperature-induced reduction in diffusion length, was determined to be 24 meV.

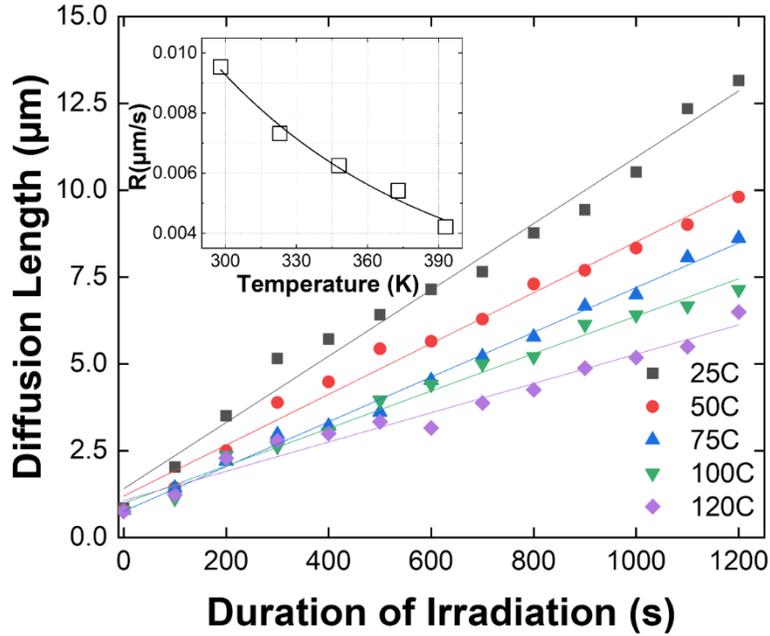

**Fig. 4.** Minority carrier diffusion length as a function of electron beam irradiation duration, measured at 25, 50, 75, 100, 120 °C. The rate of diffusion length change, $R$, for each temperature is obtained from each respective linear fit's slope. Inset depicts the relationship between $R$ and temperature and the corresponding exponential fit according to Eq. (4).

The results of the sustained beam irradiation EBIC experiments conducted at 25, 50, 75, 100, and 120 °C are presented in Fig. 4. The data acquired across the temperature range consistently show that $L$ progressively increases with the cumulative total time under electron beam irradiation. This effect was observed in various wide bandgap semiconductors such as p-GaN, unintentionally doped GaN, $p$-AlGaN, $\pi$-Ga$_2$O$_3$, n-Ga$_2$O$_3$ and $p$-ZnO [2, 8, 9, 16, 17]. The elongation of $L$ seen in Fig. 4 is characterized by its rate of change, $R$, with respect to irradiation time. The inset of Fig. 4 highlights the relationship between $R$ and temperature. The rate $R$ drops from 9.5 nm/s at 25 °C to 4.5 nm/s at 120 °C. This temperature dependence of R is given by [2]:

$$R(T) = \exp\left(\frac{\Delta E_{A,T}}{2kT}\right) \exp\left(\frac{E_{A,I}}{kT}\right) \tag{4}$$

For this equation, $R_0$ is similarly a scaling constant, $\Delta E_{A,T}$ is the same as in the Eq. (3), and $\Delta E_{A,I}$ is the activation energy associated with electron irradiation-induced process. Consequently, because Eq. (3) models the thermal dependence of $L$, it can be leveraged here to isolate and determine the activation energy of the irradiation-driven enhancement of $L$.

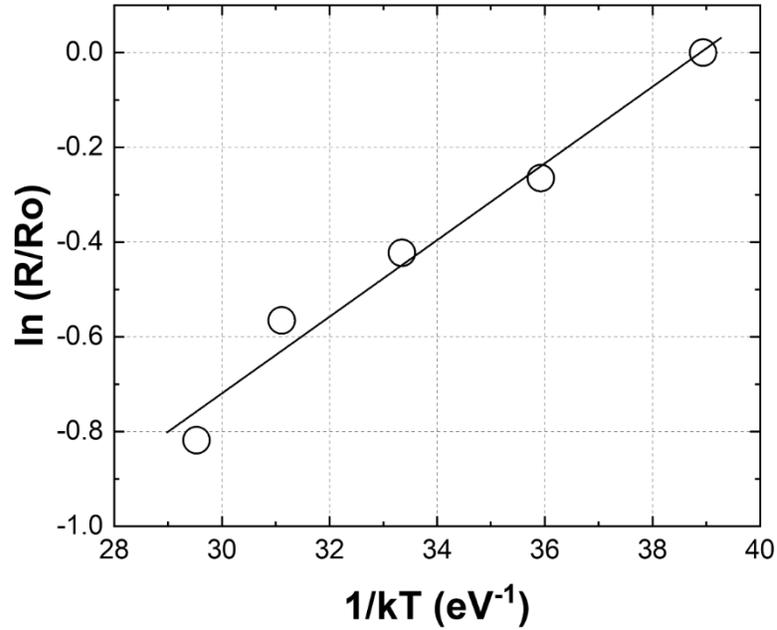

**Fig. 5.** The Arrhenius plot presenting ln ($R/R_o$) versus $1/kT$. The linear fit's slope yields an activation energy of 68 meV via Eq (4).

The Arrhenius plot presented in Fig. 5, where the slope corresponds to $\Delta E_{A,I} + 0.5\Delta E_{A,T}$, allows for the thermal and irradiation contributions to the elongation of $L$ to be separated. Utilizing this relation, $\Delta E_{A,I}$ was determined to be 68 meV, which is associated with the mechanism responsible for the elongation of $L$ resulting from prolonged electron irradiation. This activation energy agrees with Ref. [3] for undoped p-type $Ga_2O_3$ grown on the insulating Fe-doped substrate ($\Delta E_{A,I}$ = 72 meV), yet is higher than in the studies on n-type $Ga_2O_3$ which report the activation energy of 36-41 meV [9, 13]. Note that the previous report on highly resistive p-type $Ga_2O_3$ gave the activation energy of 91 meV [2]. The lower $\Delta E_{A,I}$ observed in this work for p-type $Ga_2O_3$ as compared to Ref. [2], may be indicative of a higher acceptor concentration present in the material.

An estimate for acceptor concentration ($N_A$) is obtained using the extracted activation energy of 68 meV from the EBIC analysis that is attributed to a gallium vacancy-related acceptor. The Hall measurements in Fig. 1 established a majority carrier (hole) density of about $2 \times 10^{17}$ cm$^{-3}$ at 450 K. Utilizing this for the estimate, $N_A$, was found to be $1.15 \times 10^{18}$ cm$^{-3}$ calculated from the following equation [4]:

$$p(T) = N_A \exp(-\frac{\Delta E_{A,I}}{kT}) \qquad (5)$$

Ref. [1] shows a model for the sustained irradiation effects observed. They can be summarized as follows. The energetic placement of certain deep defects within the $Ga_2O_3$ bandgap ensures they are predominately in a neutral charge state under equilibrium conditions. These deep level defect states can function as metastable traps for charges. When the material is exposed to electron beam excitation, non-equilibrium carriers are generated and consequently some of the generated excess carriers are captured by these deep defect states, which are then passivated as recombination centers, reducing the probability of non-equilibrium carrier recombination via these deep states.

This reduction in recombination pathways results in an extended carrier lifetime for the non-equilibrium charge population. The corresponding increase in $L$ is consistent with Eq. (1).

While EBIC reveals how far carriers can travel before recombination to give information about traps and defects, it doesn't distinguish between radiative and non-radiative recombination processes. To complement this technique, temperature-dependent CL measurements were carried out to obtain more information about radiative efficiency and the thermal activation energies of competing non-radiative pathways in $Ga_2O_3$. These two methods allow for a more complete picture of the minority carrier dynamics and defect landscape of the epitaxial p-$Ga_2O_3$.

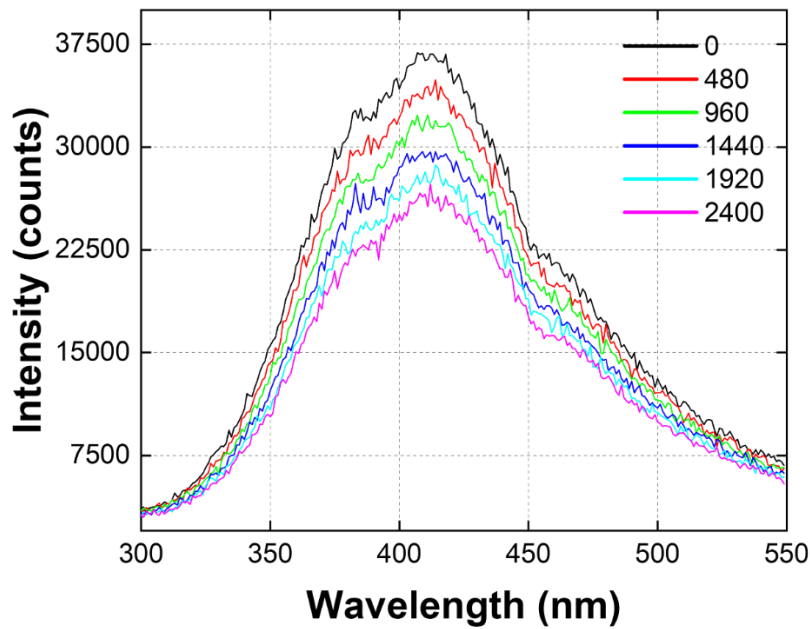

**Fig. 6.** Evolution of CL spectra for a p-Ga2O3 sample at 50 °C under prolonged electron beam irradiation. The initial spectrum (black) corresponds to the initial near zero duration irradiation and subsequent spectra were taken in 480 s intervals up to 2400 s.

Fig. 6 presents the evolution of CL spectra acquired continuously at 50 °C from a 10 μm$^2$ area on the p-type $Ga_2O_3$ epitaxial surface under electron beam irradiation exposure. The observed CL spectra are characteristic of *β*-Ga$_2$O$_3$, predominantly consisting of several broad emission bands rather than distinct near band edge emission. The broad emission bands in *β*-Ga$_2$O$_3$ are typically attributed to defect complexes involving oxygen and gallium vacancies, alongside oxygen and gallium interstitials. Several comprehensive studies of n- and p-type Ga$_2$O$_3$ optical properties have been presented in Refs. [12,14]. Fig. 6 shows a progressive decay in the overall CL intensity with increasing electron irradiation duration.

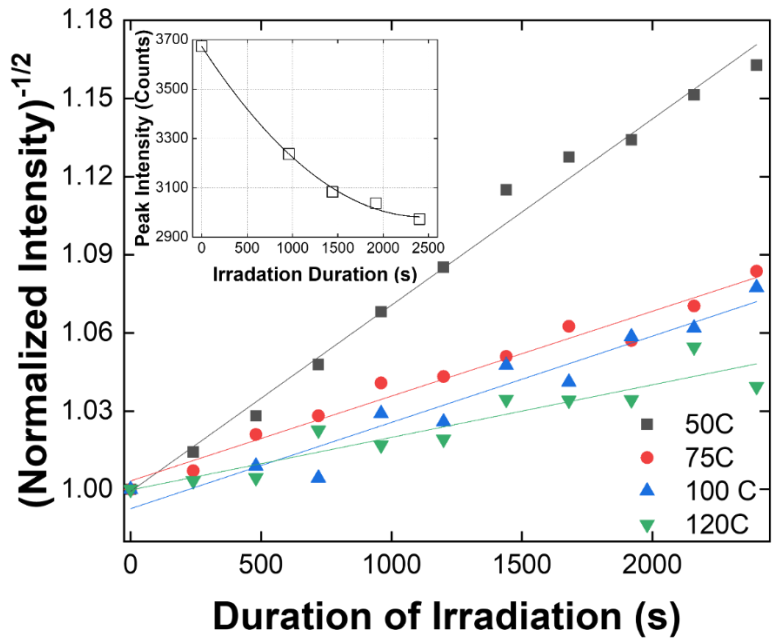

**Fig. 7.** Electron beam irradiation effects on the normalized CL peak intensity (presented as (Normalized Intensity)$^{-1/2}$) across a 50 to 120 °C range. The CL decay rates, *R*, are given by the slopes of the linear fits for each temperature. The inset illustrates the $1/t^2$ dependence of the unnormalized CL peak intensity as function of irradiation duration for the 50 °C measurement.

The peak CL intensity at 50 °C, extracted from spectra presented in Fig. 6, is plotted against electron beam irradiation duration and is presented in the inset of Fig. 7. This agrees with the previous CL report on p-type Ga$_2$O$_3$ in which both CL and EBIC were measured in the same area to demonstrate (prior to saturation effects) the near linear enhancement of *L* and corresponding CL peak intensity decrease [5]. Other reports on n-type, p-type, and highly resistive heteroepitaxially grown $Ga_2O_3$, and the mechanism responsible for sustained irradiation induced effects on minority carrier dynamics and recombination are detailed in Ref. [2, 9, 10, 14].

As Fig. 4 shows a linear increase of $L$ with increasing duration of electron irradiation, $t$, a peak intensity, $I$, (note: not the EBIC signal), is expected to be proportional to $1/t^2$, in accordance with Refs. [6, 15]. This is illustrated in the inset of Fig. 7. Considering $\tau \sim I^{-1}$ [15] and the linear dependence of $\tau^{1/2}$ on $t$ (cf. Fig.4 and Eq. (1)), one should expect a linear relationship between $I^{-1/2}$ and electron irradiation duration. This is validated and presented in Fig. 7 for temperatures ranging from 50 to 120 °C. Note that the span of CL decay diminishes at higher temperatures. The rate $R$ of CL decay for each temperature is the slope for each linear dependence. $R$ versus $T$ for the sample under investigation, shown in the inset of Fig. 8, was fitted using the following equation [2]:

$$R = R_0 \, exp\left(\frac{\Delta E_A}{2kT}\right) \qquad (6)$$

In Eq. (6), $R_0$ is the scaling constant; $\Delta E_A$ is the activation energy for CL intensity decay; $k$ is the Boltzmann constant; $T$ is temperature. The value of $\Delta E_A$ was obtained from the slope of the Arrhenius plot shown in Fig. 8 and determined to be 344 meV.

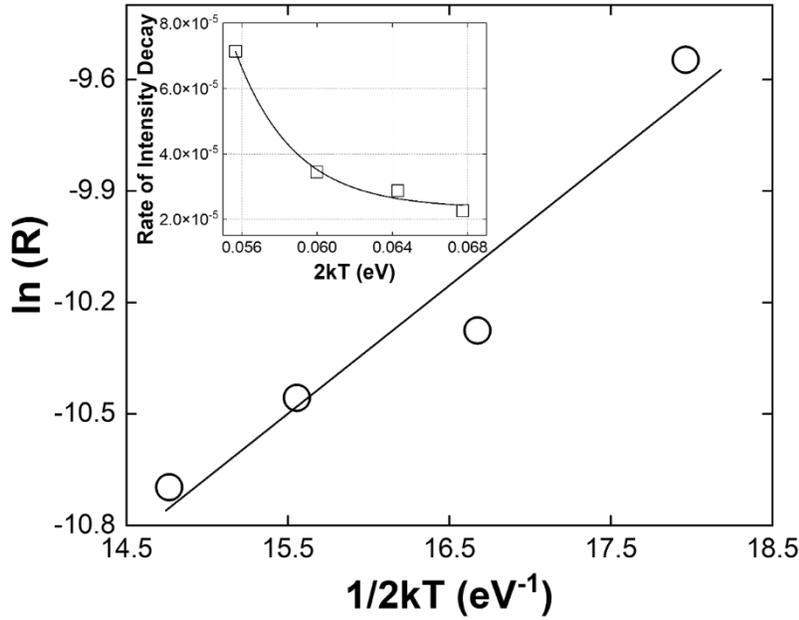

**Fig. 8.** Arrhenius plot presenting ln (*R*) vs 1/(2*kT*). The linear fit's slope yields the activation energy $\Delta E_A$. The inset depicts the relationship between *R* vs 2*kT* according to Eq. (6)

Existing literature in Ref. [17-22, 26] provides comprehensive summaries of trap states in Ga$_2$O$_3$, which are frequently linked to native crystalline imperfections and native defects like gallium and oxygen vacancies (V$_{Ga}$ and V$_O$, respectively). In Ref. [22], V$_{Ga}$ – related energetic levels are located at 100-300 meV and 300-500 meV above the top of the valence band maximum. This is in line with the CL activation energies found in this work and a prior investigation, on homoepitaxial p-type Ga$_2$O$_3$ grown on an insulating substrate, where an activation energy of 304 meV was found [5]. Similarly, the CL extracted $\Delta E_A$ of 344 meV in this work, is directly comparable to Ref. [5] and gives the same indication of deep defect levels.

**IV. Conclusions**

This investigation employed Electron Beam-Induced Current (EBIC) to characterize the minority electron diffusion length in undoped, homoepitaxial p-type Gallium Oxide grown on a conductive Sn-doped substrate. Measurements were conducted across a temperature range of 25 to 120 °C and under electron irradiation durations totaling up to 1200 s. This revealed two trends: a continuous decrease in $L$ with increasing temperature, and a near-linear increase in $L$ under prolonged irradiation duration. Both observed trends are consistent with the previously published results for p-type $Ga_2O_3$ [1-3]. The activation energy $\Delta E_{A,I}$ = 68 meV, obtained in this work, is comparable to that for p-$Ga_2O_3$ homoepitaxially grown on the Fe-doped substrate (72 meV) [3]. Based on the determined $\Delta E_{A,I}$, the gallium vacancy-related acceptor concentration, $N_A$, was calculated as $1.15 \times 10^{18}$ cm$^{-3}$.

To further elucidate the material's optoelectronic properties, temperature-dependent cathodoluminescence measurements were performed on the same undoped p-type $Ga_2O_3$ epitaxial surface. These CL studies provided insights into the radiative efficiency and the thermal activation energies of competing non-radiative pathways. An activation energy ($\Delta E_A$) of 344 meV, associated with the thermal quenching of CL intensity was determined. This $\Delta E_A$ aligns well with gallium vacancy-related defect levels within the 300 to 500 meV range reported in Ref. [22], as well as agreeing with the previous study on homoepitaxial p-type $Ga_2O_3$ grown on an insulating substrate [5]. The presence and characteristics of these defects, as evidenced by their distinct impacts on both carrier diffusion length and radiative efficiency, are critical considerations for the future development and optimization of $Ga_2O_3$-based electronic and optoelectronic devices.

**V. Acknowledgements**

The research at UCF was supported in part by NSF (Grant Nos. ECCS2310285, ECCS2341747, and ECCS 2427262), US–Israel BSF (Award No. 2022056), and NATO (Award Nos. G6072 and G6194). The work at UF was performed as part of the Interaction of Ionizing Radiation with Matter University Research Alliance (IIRM-URA), sponsored by the Department of the Defense, Defense Threat Reduction Agency, under Award No. HDTRA1-20-2-0002, monitored by Jacob Calkins. The present work is a part of the "GALLIA" International Research Project, CNRS, France. GEMaC colleagues acknowledge financial support of French National Agency of Research (ANR), project "GOPOWER," Grant No. CE-50 N0015-01. French and Israeli researchers acknowledge the collaborative PHC-Maimonide project N50047TD. The research at Tel Aviv University was partially supported by the US–Israel BSF (Award No. 2022056) and NATO (Award No. G6072).